\documentclass[runningheads]{llncs}
\usepackage[T1]{fontenc}
\usepackage{graphicx}
\usepackage{multirow}
\usepackage[outdir=./]{epstopdf}
\usepackage{subcaption}
\usepackage{caption} 
\begin{document}
%
\title{NCIS: Deep Color Gradient Maps Regression and Three-Class Pixel Classification for Enhanced Neuronal Cell Instance Segmentation in Nissl-Stained Histological Images}


%
\titlerunning{NCIS}
%

\author{Valentina Vadori\inst{1} \and
Antonella Peruffo\inst{2} \and
Jean-Marie Graïc\inst{2} \and
Livio Finos\inst{3} \and
Livio Corain\inst{4} \and
Enrico Grisan\inst{1}}

\authorrunning{Vadori et al.}
%

\institute{Dept. of Computer Science \& Informatics, London South Bank University, UK\\
\email{\{vvadori,egrisan\}@lsbu.ac.uk} \and
Dept. of Comparative Biomedicine \& Food Science, University of Padova, IT\\
\email{\{antonella.peruffo,jeanmarie.graic\}@unipd.it} \and
Dept. of Developmental Psychology and Socialisation, University of Padova, IT \\
\email{livio.finos@unipd.it}\and
Dept. of Management and Engineering, University of Padova, IT\\
\email{livio.corain@unipd.it}}


%
\maketitle              
\begin{abstract}
Deep learning has proven to be more effective than other methods in medical image analysis, including the seemingly simple but challenging task of segmenting individual cells, an essential step for many biological studies. Comparative neuroanatomy studies are an example where the instance segmentation of neuronal cells is crucial for cytoarchitecture characterization. This paper presents an end-to-end framework to automatically segment single neuronal cells in Nissl-stained histological images of the brain, thus aiming to enable solid morphological and structural analyses for the investigation of changes in the brain cytoarchitecture. A U-Net-like architecture with an EfficientNet as the encoder and two decoding branches is exploited to regress four color gradient maps and classify pixels into contours between touching cells, cell bodies, or background. The decoding branches are connected through attention gates to share relevant features, and their outputs are combined to return the instance segmentation of the cells. The method was tested on images of the cerebral cortex and cerebellum, outperforming other recent deep-learning-based approaches for the instance segmentation of cells.

\end{abstract}
\section{Introduction}
Advancements in microscopy have made it possible to capture \textit{Whole Slide Images} (WSIs) and obtain cellular-level details, revealing the intricate nature of the brain cytoarchitecture. This progress has opened up new avenues for conducting quantitative analysis of cell populations, their distribution, and morphology, which can help us answer a range of biological questions. Comparative neuroanatomy studies examine differences in brain anatomy between groups distinguished by factors like sex, age, pathology, or species, investigating the connections between the brain's structure and function \cite{AMUNTS20071061,graic2022primary,corain2020multi}. 
A standard analysis pipeline involves the use of Nissl stain to label neuronal cells in tissue sections (histological slices) of brain specimens \cite{garcia2016distinction}. These sections are then fixed and digitized as WSIs for examination. WSIs are often characterized by their sheer size and complexity, making computerized methods necessary for their efficient and reproducible processing.
Automatic cell instance segmentation plays a crucial role, as it allows to extract features at the single cell level.

In the field of digital pathology, numerous methods have been proposed to segment cells and nuclei and aid the detection and diagnosis of diseases. 
These methods mainly rely on a set of algorithms, including intensity thresholding, morphology operations, watershed transform, deformable models, clustering, graph-based approaches \cite{XingRobust}. However, cell instance segmentation is very challenging due to the varying size, density, intensity and texture of cells in different anatomical regions, with additional artifacts that can easily influence the results. 
Recently, deep learning has shown remarkable progress in medical image analysis, and neural networks have been successfully applied for cell segmentation, achieving higher quality than traditional algorithms. In the last years, a considerable number of approaches have adopted a semantic segmentation formulation that employs a U-net-like convolutional neural network architecture \cite{ronneberger2015u}. These methods incorporate customized post-processing techniques, such as marker-controlled watershed algorithms and morphological operations, to separate cells instances. Some integrate the formulation with a regression task. Huaqian et. al \cite{wu2022general} propose a framework with an EfficientNet as the U-Net encoder for ternary classification (contours between touching cells, cell bodies, and \textit{background}, BG). Ultimate erosion and dynamic dilation reconstruction are used to determine the markers for watershed. 
StarDist \cite{schmidt2018cell} regresses a star-convex polygon for every pixel.
CIA-Net \cite{zhou2019cia} exploits two decoders, where each decoder segments either the nuclei or the contours.
Hover-Net \cite{graham2019hover} uses a Preact-ResNet50 based encoder and three decoders for \textit{foreground} (FG)/BG segmentation, cell type segmentation, and regression of horizontal and vertical distances of pixels from the cell centroid.  
Mesmer \cite{greenwald2022whole} considers classification into whole contours, cell interiors, or BG, and regression of the distance from the cell centroid.

Since in histological images the boundaries between cells that are in contact are often incomplete or ambiguous and they can appear between cells with differing characteristics and orientations, we recognized the pivotal role of correctly predicting these boundaries for accurate cell separation. Therefore, we have developed an approach that focuses on enhancing the prediction of  contours. Specifically, we propose NCIS as an end-to-end framework to automatically segment individual neuronal cells in Nissl-stained histological images of the brain. NCIS employs an U-Net-like architecture, which synergistically combines solutions from \cite{wu2022general,graham2019hover,zhou2019cia} to classify pixels as contours between touching cells, cell body, or BG, and to regress four color gradient maps that represent distances of pixels from the cell centroid and that are post-processed to get a binary mask of contours. Since cells are often slanted and there are configurations where smaller cells burrow into the concavities of larger ones, we hypothesized that the prediction of diagonal gradients together with horizontal and vertical ones could help to strengthen the approach. NCIS was created to examine the cytoarchitecture of brain regions in diverse animals, including cetaceans, primates, and ungulates. The primary objective is to conduct comparative neuroanatomy studies, with a particular emphasis on diseases that impair brain structure and functionality, such as neurodegeneration and neuroinflammation. We tested NCIS on images of the auditory cortex and cerebellum, outperforming other recent deep-learning-based approaches for the instance segmentation of cells. 
\section{Dataset}
The data of this study are a set of $53$ $2048$x$2048$ histological images extracted from Nissl-stained $40$x magnification WSIs of the auditory cortex of Tursiops truncatus, also known as the bottlenose dolphin. Brain tissues were sampled from $20$ specimens of different subjects (new-born, adult, old) stored in the Mediterranean Marine Mammals Tissue Bank (http://www.marinemammals.eu) at the University of Padova, a CITES recognized (IT020) research center and tissue bank. These specimens originated from stranded cetaceans with a decomposition and conservation code (DCC) of $1$ and $2$, which align with the guidelines for post-mortem investigation of cetaceans \cite{ijsseldijk2019best}.
The images were divided into $3$ subsets: $42$ images for training, $5$ for validation and $6$ for testing. 

To assess the generalizability of the proposed method on cerebral areas not seen during training, we considered an independent $2048$x$2048$ Nissl-stained histological image of the cerebellum of the bovine \cite{corain2020multi}. In this case, the animal was treated according to the present European Community Council directive concerning animal welfare during the commercial slaughtering process and was constantly monitored under mandatory official veterinary medical care. 
Compared to the training set, the additional image is characterized by the presence of a higher density granular layer of touching or overlapping cells with small dimensions and predominantly circular shape and a thin Purkinje cell layer with relatively larger and sparse pear-shaped cells.

All images were annotated using QuPath \cite{bankhead2017qupath} software, resulting in $24,044$ annotated cells of the auditory cortex, and $3,706$ of the cerebellum.


\section{Method}
\begin{figure}
\includegraphics[width=\textwidth]{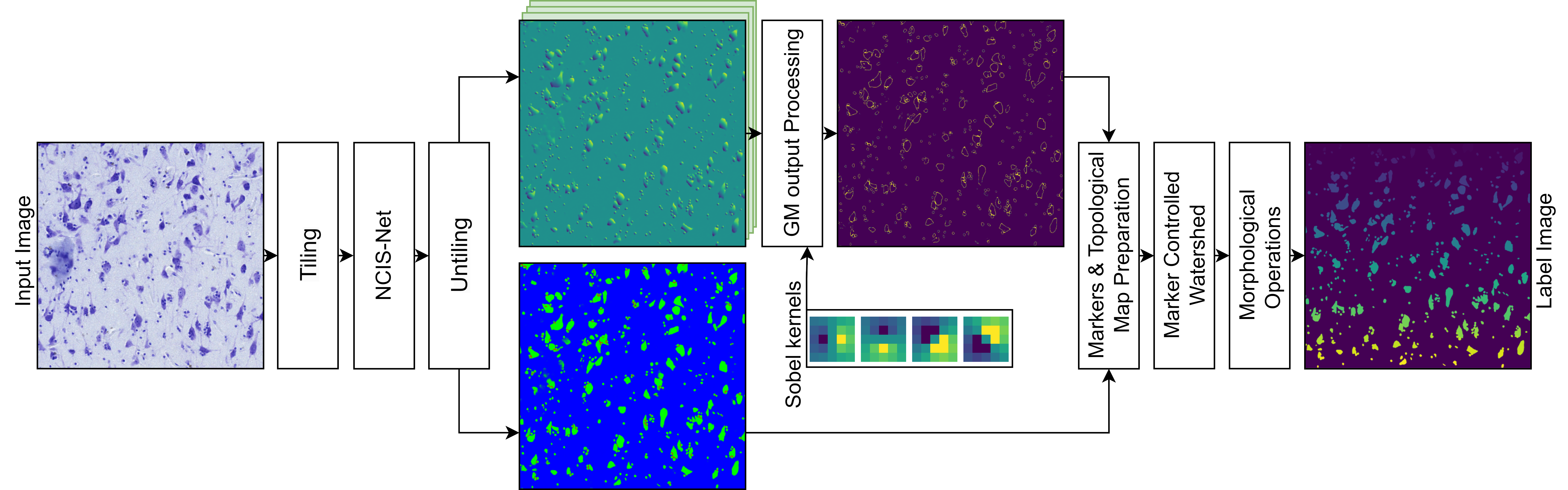}
\caption{An overview of the overall NCIS approach for neuronal cell instance segmentation in Nissl-stained whole slice images.} \label{NCISdiagram}
\end{figure}
\begin{figure}
\includegraphics[width=\textwidth]{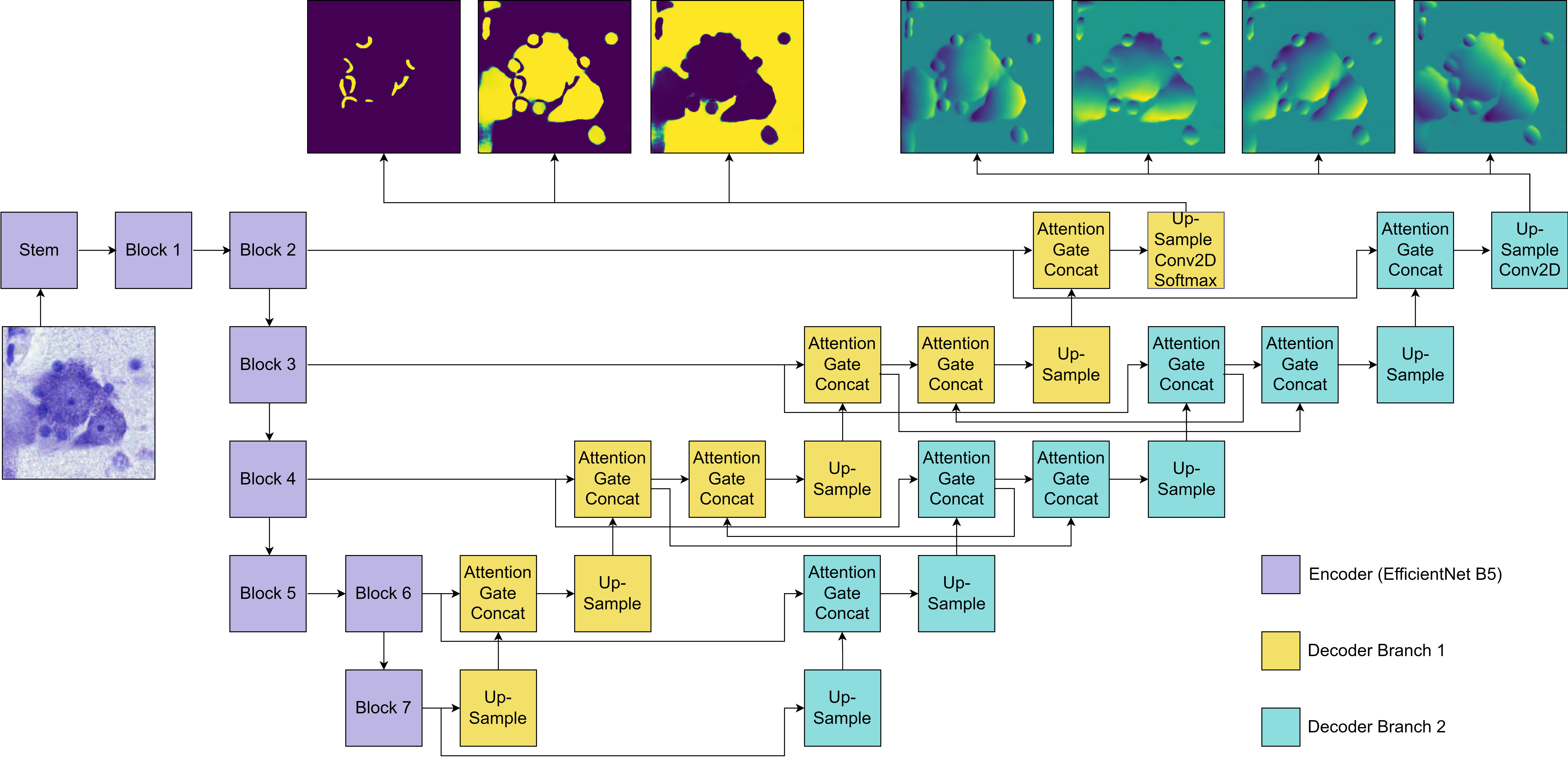}
\caption{An overview of the architecture of the proposed NCIS-Net.} \label{NCISarchitecture}
\end{figure}

The proposed NCIS framework for automatic instance segmentation of neuronal cells can be observed in Fig. \ref{NCISdiagram}.
An image of arbitrary size is divided into patches of size $256$x$256$ via a tiling step with $50\%$ overlap between patches. Patches are individually fed to the deep learning model, NCIS-Net, whose architecture is shown in Fig. \ref{NCISarchitecture}. The model outputs for each patch are combined via a untiling step to get $7$ outputs with size equal to that of the original image. Finally, a series of post-processing steps is applied to generate the instance segmentation. 
\subsection{NCIS-Net} 
\subsubsection{Architecture} The proposed NCIS-Net has a U-Net-like structure, as shown in Fig. \ref{NCISarchitecture}. The down-sampling, feature extracting branch of the encoder is based upon a state-of-the-art network, \textit{EfficientNet-B5}, whose building blocks are arrays of mobile inverted bottleneck \textit{MBConv} optimized with neural architecture search \cite{tan2019efficientnet}. 
NCIS-Net is characterized by two decoding branches for a classification and regression task. The first branch performs three-class pixel classification. Its output is a three-channel image with probabilities for boundaries between touching cells, cell bodies, and BG, respectively. The union of the first two classes constitutes the FG. The second branch regresses four color gradient maps. Its output is a four-channel image, where each channel represents a color gradient map with vertical, horizontal, and diagonal directions. As illustrated in Fig. \ref{NCISarchitecture}, in a gradient map, each individual cell is represented by a color gradient where pixel values increment from the minimum of $-1$ (blue in the figure) to the maximum of $1$ (yellow in the figure) according to the prescribed direction (vertical from left to right, horizontal from top to bottom, diagonal from bottom left to top right and diagonal from top left to bottom right). $0$ corresponds to the cell centroid. Four skip connections leveraging attention gates \cite{oktay2018attention} connect the encoder branch to each decoder branch, promoting a focused exploitation of the features encoded at different resolutions. Skip connections with attention gates are also introduced between the two decoder branches to favor feature sharing. 
\subsubsection{Loss function}
NCIS-Net is trained through the backpropagation algorithm applied to a loss function $\mathcal{L} = \mathcal{L}_{PC}+\mathcal{L}_{GR}$ that jointly optimizes the encoder, pixel classification ($PC$) and gradient regression ($GR$) branches, with
\begin{equation}
\mathcal{L}_{PC} = \lambda_1\mathcal{C}+\lambda_2\mathcal{D}_1+\lambda_3\mathcal{D}_2, \ \ \ \ \ \ \mathcal{L}_{GR} = \lambda_4\mathcal{M}_1+\lambda_5\mathcal{M}_2
\label{pcloss}
\end{equation}
%
where $\lambda_1 = \lambda_3 = \lambda_4 = \lambda_5 = 2, \lambda_2 = 1$ are weighting factors set via hyperparameter validation. $\mathcal{C}$ is the categorical cross-entropy between the ground truth (GT) and $\hat{Y}_{PC}$, the output of the $PC$ branch.
$\mathcal{D}_k$ is the Dice loss for class $k$:
\begin{equation}
\mathcal{D}_k = 1-\frac{2\sum_{i=1}^{N} (Y_{PC,i,k}\hat{Y}_{PC,i,k})+\epsilon       }{\sum_{i=1}^{N} Y_{PC,i,k}+\sum_{i=1}^{N} \hat{Y}_{PC,i,k}+\epsilon}
\end{equation}
where $N$ is the total number of pixels in the input image, $\epsilon $ is a smoothness constant. Specifically, $\mathcal{D}_1$ and $\mathcal{D}_2$ in Eq. \ref{pcloss} are the Dice losses for contours and cell bodies, respectively. 
$\mathcal{M}$ indicates a mean squared error loss. $\mathcal{M}_1$ is the mean squared error loss between the GT ($Y_{GR}$) and the predicted output $\hat{Y}_{GR}$ of the $GR$ branch, while $\mathcal{M}_2$ is defined as:
 \begin{equation}
\mathcal{M}_2 = \frac{1}{N\cdot J}\sum_{i=1}^{N}\sum_{j=1}^{J}
(\nabla_j Y_{GR,i,j}-\nabla_j\hat{Y}_{GR,i,j})^2
\end{equation}
where $J$ is the number of gradient maps ($4$ in our case) and $\nabla$ is the derivative operator. Note that for each gradient map, finite derivatives are taken by convolving the map with $5x5$ Sobel kernels with orientation and direction matching that of the corresponding gradient map, as illustrated in Fig. \ref{NCISdiagram}.

\subsubsection{Training} Data batches are created from the training set of $2048$x$2048$ images. Images are picked randomly with replacement and a series of random augmentations including rotations, flipping, deformations, intensity variations, and blurring, are applied to each selected image and corresponding three-class semantic and cell instance segmentation GT masks, when necessary. GT gradient maps are computed at this stage based on the transformed instance segmentation GT masks. Random cropping is then applied to the transformed images to get $256$x$256$ images to be fed to the network. By utilizing this approach, it is highly likely that the batches used during each epoch will not only have varying transformations but also distinct cropped original images, ultimately aiding in the prevention of overfitting. All models are trained with the TensorFlow $2$ framework in two phases. In the first phase the encoder, pre-trained on the \textit{Image-Net} dataset \cite{deng2009imagenet}, is frozen, while in the second fine tuning phase the encoder is unfrozen, with the exception of batch normalization layers. Each phase continues for a maximum of $50$ epochs with early stopping (patience of $8$ epochs, $400$ batches per epoch, batch size of $16$). The validation set for early stopping is created from the validation set of $2048$x$2048$ images by cropping them into $256$x$256$ with no overlapping and no augmentations. We use the AMSGrad optimizer with clipnorm of $0.001$ and a learning rate of $10^{-3}$ in the first phase and $10^{-5}$ in the second phase.
The deep learning model architecture, training and inference code will be made available upon acceptance on GitHub.
\subsection{Untiling and Post-Processing}
Patches from the same image are blended together via interpolation with a second order spline window function that weights pixels when merging patches \cite{guillaume2017}. 

Within each of the color gradient map, pixels between different cells should have a meaningful difference. Therefore, Sobel kernels are utilized to get the following contour map: 
\begin{equation}
C_{PC,i} = max(\nabla_1\hat{Y}_{GR,i,1},\nabla_2\hat{Y}_{GR,i,2},\nabla_3\hat{Y}_{GR,i,3},\nabla_4\hat{Y}_{GR,i,4})
\end{equation}
where $\hat{Y}_{GR,i,j}, j= 1,...,4,$ is normalized between $0$ and $1$. $C_{PC,i}$ is thresholded on a per-image basis via the triangle method \cite{zack1977automatic}, so that ones in the thresholded binary version $C_{PC,i,th}$ correspond to contours. 
A binary mask is defined with pixels set to $1$ if they are most likely to belong to cell bodies based on the outputs of both decoders ($\hat{Y}_{PC,i,2}$>$\hat{Y}_{PC,i,1}$ and $\hat{Y}_{PC,i,2}$>$\hat{Y}_{PC,i,3}$ and $C_{PC,i,th} = 0$). Connected components smaller than $80$ pixels are removed. The mask is eroded with a disk-shaped structuring element of radius $4$ to force the separation of touching cells that are only partially separated. The resulting connected components bigger than $3$ pixels are used as markers for the marker-controlled watershed algorithm applied to the topological map given by the complement of $\hat{Y}_{PC,i,2}$. A foreground binary mask is obtained as $\hat{Y}_{PC,i,1}$+$\hat{Y}_{PC,i,2}$>$\hat{Y}_{PC,i,3}$. After holes filling and  morphological opening to smoothen the cell boundaries, it is used to constrain the watershed labelling. Segmentations are refined via morphological closing and opening, ensuring that boundaries between cells are maintained. 

\begin{table*}
\centering
\caption{Segmentation performance of the proposed NCIS method and compared approaches.  \#P indicates the number of network trainable parameters.}
\label{tabresults}
\
\begin{tabular}{ll|cc|cc}
\hline
\multirow{2}{*}{\textbf{Methods}} &              & \multicolumn{2}{l|}{\textbf{Auditory Cortex}} & \multicolumn{2}{l}{\textbf{Cerebellum}} \\ \cline{3-6} 
                                  & \#P(M) & Dice                 & AP@0.5        & Dice                   & AP@0.5        \\ \hline
Huaqian et. al \cite{wu2022general}                       & $40.15$             & $0.915$              & $0.782$         & $0.708$                 &$0.329$         \\
Hover-Net \cite{graham2019hover}                        & $44.98$             & $0.912$          & $0.780$         & $0.792$              & $0.230$         \\
Mesmer \cite{greenwald2022whole}                                            &    $25.50$          & $0.862$             & $0.657$         & $\mathbf{0.801}$            & $0.156$         \\ \hline
NCIS - \textit{no attention}       &  $58.07$            & $0.920$                & $0.810$          & $\mathbf{0.801}$               & $\mathbf{0.458}$         \\
NCIS                              & $61.91$             & $\mathbf{0.925}$               & $\mathbf{0.814}$         & $0.768	$             & $0.402$         \\ \hline
\end{tabular}
\end{table*}
\captionsetup[subfigure]{labelformat=empty}

\section{Results}

\subsection{Evaluation Metrics}
To evaluate the semantic segmentation performance on test images, we utilize the Dice Coefficient (Dice). Instance segmentation performance is evaluated according to the average precision with threshold $0.5$ (AP@0.5). Predicted instances are compared to the GT and a match (\textit{true positive}, TP) is established if a GT object exists whose intersection over union (IoU) is greater than $0.5$. Unmatched objects are counted as \textit{false positive} (FP) and unmatched GT objects are \textit{false negatives} (FN). The average precision is then given by AP = TP/(TP+FP+FN). 
\subsection{Experimental Results}

We evaluated NCIS, Huaqian et. al \cite{wu2022general}, Hover-Net \cite{graham2019hover}, and Mesmer \cite{greenwald2022whole} on images of the auditory cortex. The Dice coefficient for all methods (except Mesmer) is higher than $0.912$, as shown in Table \ref{tabresults}, highlighting that U-net-like architectures can produce reliable results for semantic segmentation regardless of the specific architecture. Huaqian et. al, based on three-class pixel classification, and Hover-Net, based on binary pixel classification and regression of two distances from the cell centroid, achieve similar performance. NCIS, which focuses on contours predictions through three-class pixel classification and regression of four distances (or color gradients) from the cell centroid, displays the best performance in terms of both semantic and instance segmentation accuracy. NCIS - \textit{no attention}, the NCIS version with no attention gates, also performs well compared to the other methods but slightly worse than NCIS. Mesmer, which regresses the distance from the cell centroid to determine markers instead of contours during the post-processing, performs substantially worse.
 
When testing the methods on the cerebellum, an area not seen during training, we can observe lower performance overall, as expected. Interestingly, NCIS - \textit{no attention} is the overall top performer, indicating that attention may be detrimental if the training set does not adequately represent the test set. For semantic segmentation, Mesmer is on par with NCIS - \textit{no attention}, followed by Hover-Net. NCIS is the second-best for instance segmentation.

\begin{figure}%
\centering
\begin{subfigure}{.198\columnwidth}
\includegraphics[width=\columnwidth]{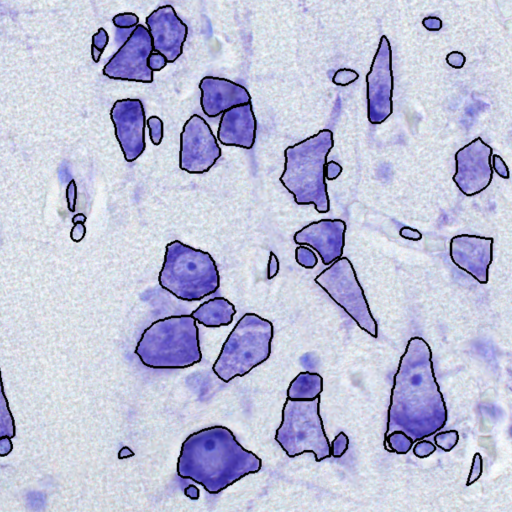}%
\end{subfigure}\hfill%
\begin{subfigure}{.198\columnwidth}
\includegraphics[width=\columnwidth]{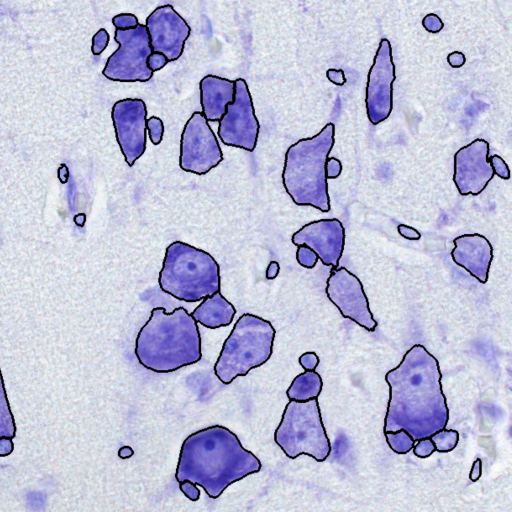}%
\end{subfigure}\hfill%
\begin{subfigure}{.198\columnwidth}
\includegraphics[width=\columnwidth]{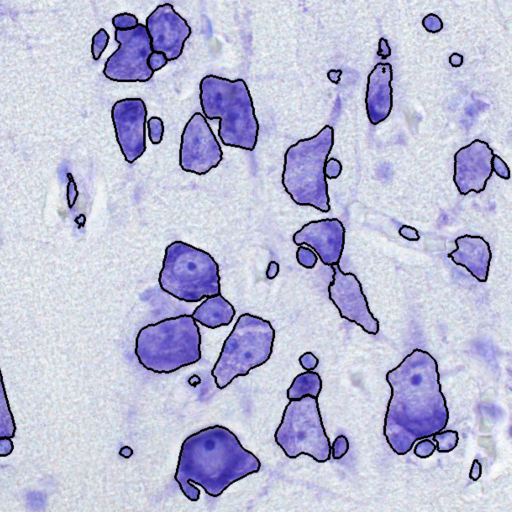}%
\end{subfigure}\hfill%
\begin{subfigure}{.198\columnwidth}
\includegraphics[width=\columnwidth]{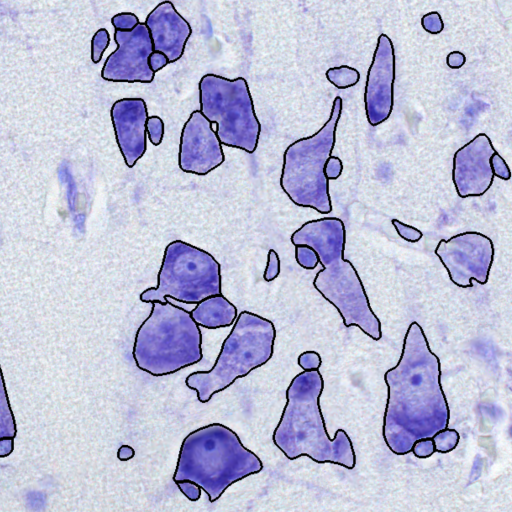}%
\end{subfigure}\hfill%
\begin{subfigure}{.198\columnwidth}
\includegraphics[width=\columnwidth]{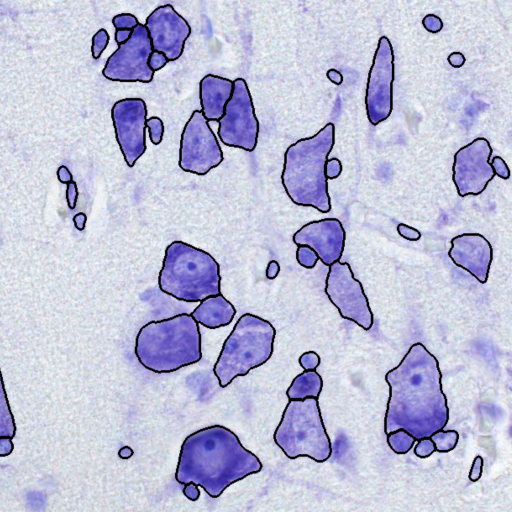}%
\end{subfigure}\hfill%
\par
\begin{subfigure}{.198\columnwidth}
\includegraphics[width=\columnwidth]{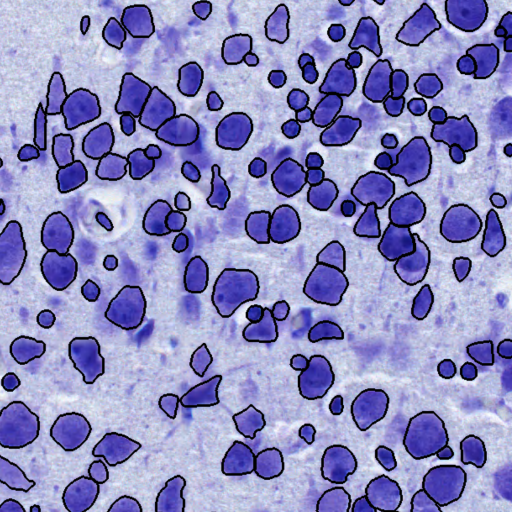}%
\end{subfigure}\hfill%
\begin{subfigure}{.198\columnwidth}
\includegraphics[width=\columnwidth]{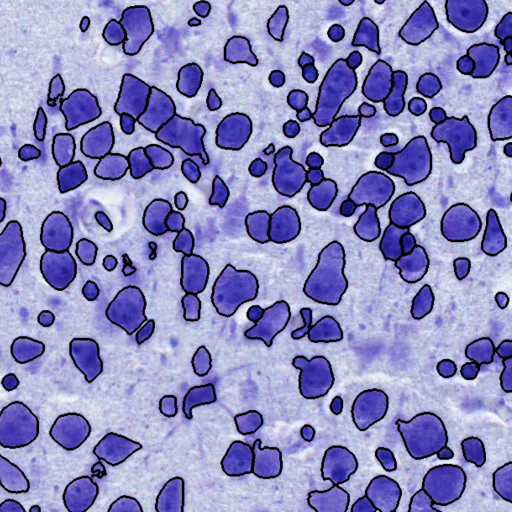}%
\end{subfigure}\hfill%
\begin{subfigure}{.198\columnwidth}
\includegraphics[width=\columnwidth]{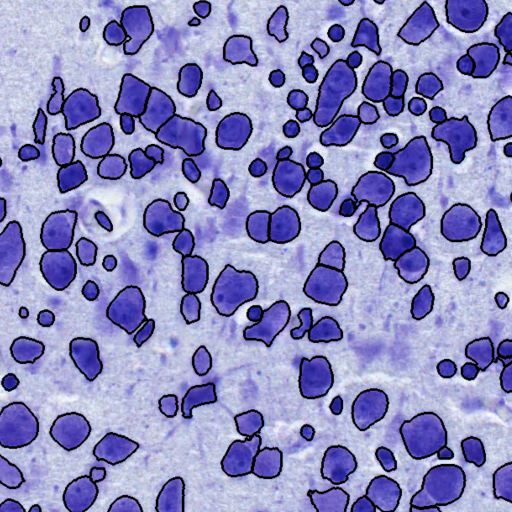}%
\end{subfigure}\hfill%
\begin{subfigure}{.198\columnwidth}
\includegraphics[width=\columnwidth]{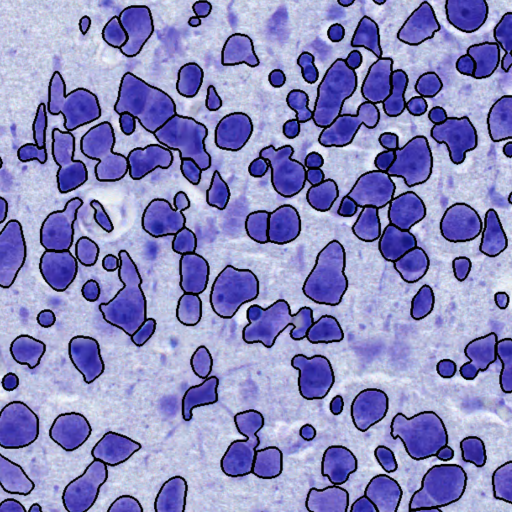}%
\end{subfigure}\hfill%
\begin{subfigure}{.198\columnwidth}
\includegraphics[width=\columnwidth]{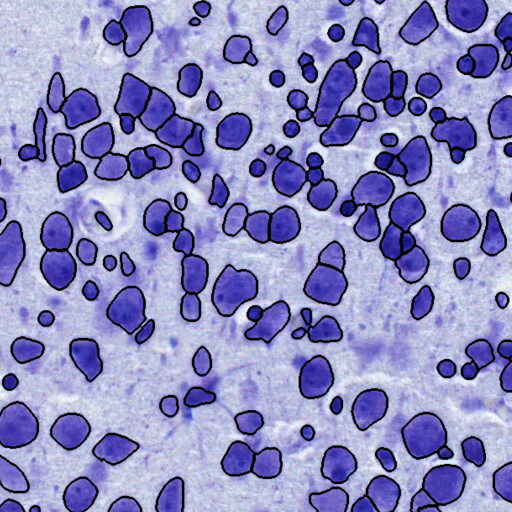}%
\end{subfigure}\hfill%
\par
\begin{subfigure}{.198\columnwidth}
\includegraphics[width=\columnwidth]{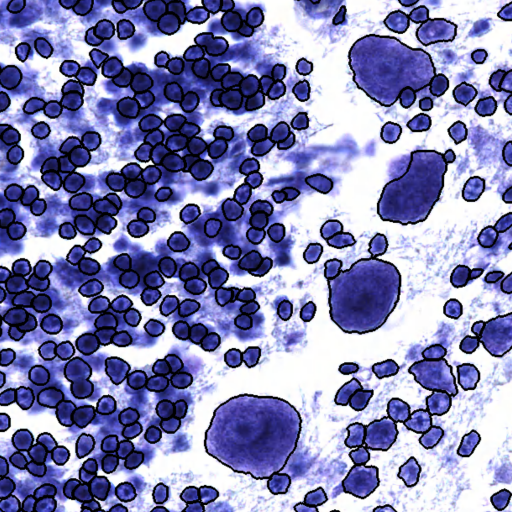}%
\caption{GT}%
\end{subfigure}\hfill%
\begin{subfigure}{.198\columnwidth}
\includegraphics[width=\columnwidth]{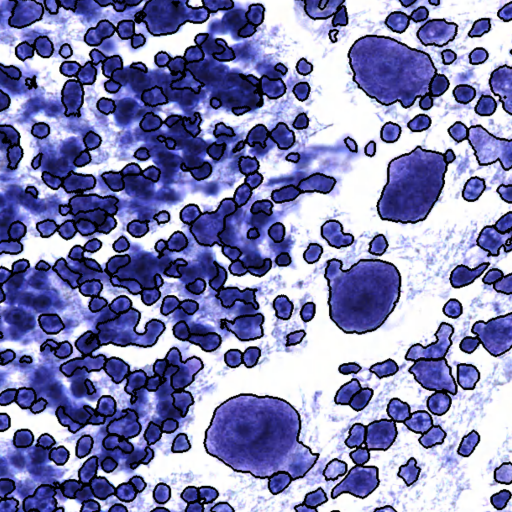}%
\caption{Huaqian et. al}%
\end{subfigure}\hfill%
\begin{subfigure}{.198\columnwidth}
\includegraphics[width=\columnwidth]{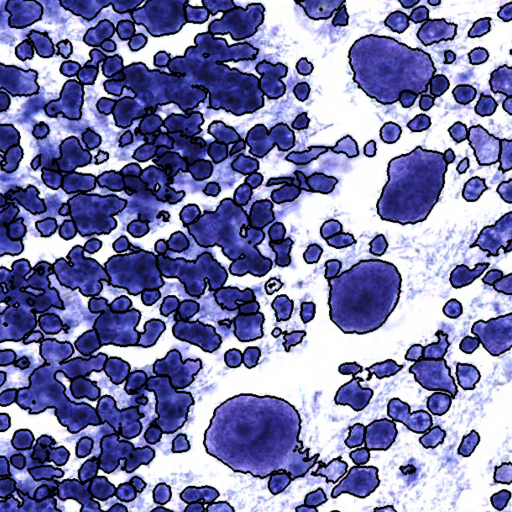}%
\caption{Hover-Net}%
\end{subfigure}\hfill%
\begin{subfigure}{.198\columnwidth}
\includegraphics[width=\columnwidth]{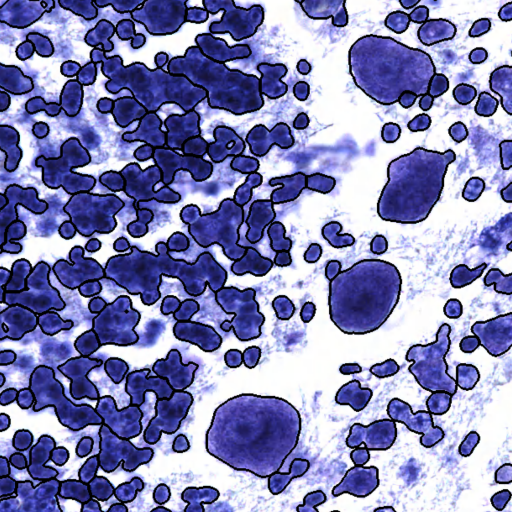}%
\caption{Mesmer}%
\end{subfigure}\hfill%
\begin{subfigure}{.198\columnwidth}
\includegraphics[width=\columnwidth]{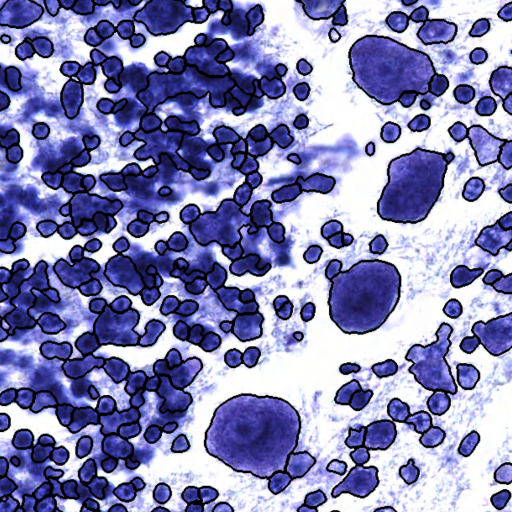}%
\caption{NCIS}%
\end{subfigure}\hfill%
\caption{Qualitative results for two sample tiles from the Auditory Cortex (top and center) and Cerebellum (bottom) datasets. GT is shown in the first column.}
\label{figabc}
\end{figure}

The qualitative outcomes for three example tiles are presented in Fig. \ref{figabc}. The segmentation of NCIS appears to be visually appealing, being smoother and more conforming to ground truth. Common segmentation errors include cell merging and inaccurate identification of artifacts as cells. 
For the cerebellum, all approaches struggle on the higher density granular layer, but the Purkinje cells, which closely match the cells encountered during training, are correctly isolated.

\section{Discussion}

Currently, there are limited techniques tailored for the segmentation of neuronal cells in Nissl-stained histological slices of the brain. To address this issue, we propose a new segmentation framework, called NCIS, which employs a dual-decoder U-Net architecture to enhance contour prediction by combining three-class pixel classification and regression of four color gradient maps. Our model outperforms existing state-of-the-art methods on images of the auditory cortex, demonstrating its ability to effectively deal with the challenges of neuronal cell segmentation (cells with variable shapes, intensity and texture, possibly touching or overlapping). If tested on an area of the brain not seen during training, the NCIS semantic segmentation accuracy is promising, but the instance segmentation performance indicates the need to enrich the training set. We believe that the number of NCIS-Net parameters could be reduced without compromising performance through minor architectural changes (e.g., summation instead of concatenation in skip connections). 

NCIS could be particularly useful in processing histological WSIs of different species for comparative neuroanatomy studies, potentially contributing to the understanding of neurodegenerative and neuroinflammatory disorders.


%
%
 \bibliographystyle{splncs04}
 \bibliography{refs}
%
%
%
%
%
\end{document}